\documentclass[twocolumn,showpacs,preprintnumbers,amsmath,amssymb]{revtex4}
\usepackage{graphicx}
\usepackage{bm}
\usepackage{dcolumn}
\usepackage{hyperref}

\def\sci#1#2#3{{Science} {\bf #1}, #2 (#3)}
\def\prl#1#2#3{{ Phys.   Rev.   Lett.  } {\bf #1}, #2 (#3)}
\def\pla#1#2#3{Phys.   Lett.   A {\bf #1}, #2 (#3)}
\def\pra#1#2#3{Phys.   Rev.   A {\bf #1}, #2 (#3)}

\def\rmp#1#2#3{Rev.   Mod.   Phys. {\bf #1}, #2 (#3)}
\def\nl#1#2#3{Nature(London) {\bf #1}, #2 (#3)}

\def\jpb#1#2#3{J. Phys. B  {\bf #1}, #2 (#3)}

\def\noi{\noindent}
\def\bc{\begin{center}}
\def\ec{\end{center}}
\newcommand{\bea}{\begin{equation}}
\newcommand{\eea}{\end{equation}\noi}
\newcommand{\ber}{\begin{eqnarray}}
\newcommand{\eer}{\end{eqnarray}\noi}
\begin{document}
\title{Finite temperature scaling theory for the collapse of Bose-Einstein condensate}
\author{Shyamal Biswas}\email{tpsb2@iacs.res.in}
\affiliation{Department of Theoretical Physics, Indian Association for the Cultivation of Science, Jadavpur, Kolkata 700032, INDIA}
\date{\today}
\begin{abstract}
We show how to apply the scaling theory in an inhomogeneous system
like harmonically trapped Bose condensate at finite temperature. We
calculate the temperature dependence of the critical number of
particles by a scaling theory within the Hartree-Fock approximation
and find that there is a dramatic increase in the critical number of
particles as the condensation point is approached.
\end{abstract}
\maketitle
\section{Introduction}
Experimental\cite{1,2,3,4,5,6,7} and theoretical\cite{8,9,10,11,12} studies of Bose-Einstein condensation(BEC) reveal the fact that Bose-Einstein condensates of trapped ultracold atomic gases
are influenced by atomic interaction\cite{8,9,10,12,13,14} which in general is characterized by s-wave scattering
length($a_s$) and can be tuned\cite{15,16,17,18} by an external magnetic field. Stability and
collapse of the Bose gas with negative scattering length has been
observed in the clouds of ultracold $^7$Li\cite{17,18} and
$^{85}$Rb\cite{19,20}. If the interaction is attractive, the gas
tends to increase the density of the central region of the trap.
This tendency is opposed by the zero-point energy and thermal energy
of the atoms. If the number of atoms is greater than a critical
number($N_c$), the central density increases strongly and the
zero-point and thermal energies are no longer able to avoid the
collapse of the gas.

Let us first consider a many particle system of 3-D isotropic
harmonically trapped ideal Bose gas of $N$ alkali atoms. The system
is in equilibrium with its surroundings at temperature($T$). Let the
mass of each particle be $m$. The ground state of a single particle
is macroscopically populated below a certain temperature which is
called the Bose-Einstein condensation temperature and is given by
\cite{8} $T_c=\frac{\hbar\omega}{k}[\frac{N}{\zeta(3)}]^{1/3}$,
where $\omega$ is the angular frequency of the harmonic oscillators
and $k$ is the Boltzmann constant. Now we consider the gas to be
interacting weakly and attractively by the two body interacting
potential $V_{int}({\bf{r}})=g\delta^3({\bf{r}})$, where
$g=-\frac{4\pi\hbar^2a}{m}$ is the coupling constant and $a=-a_s$ is
the absolute value of the s-wave scattering length. For $0\le T<T_c$ the length scale of the system becomes
$l=\sqrt{\frac{\hbar}{m\omega}}$, which is the length scale of the
ground state of the harmonic oscillators. The typical two body
interaction energy for $N$ number of particles is\cite{8} $\sim N^2
g/2V$, where $V$ is the volume of the system. For the attractive
interaction, the gas tends to increase the density of the central
region of the trap. For $T=0$, this tendency is resisted by the
zero-point energy of the atoms. If the number of atoms is greater
than a critical number, the central density increases so much that
the zero-point energy is no longer able to avoid the collapse of the
gas. In the critical situation, the typical oscillator energy must
be comparable to the typical interaction energy and as a consequence
we have $\frac{N_ca}{l}\sim 1$.

However, for $0<T<T_c$, although length scale of the system is $l$ yet the thermal fluctuation dominates over
the quantum fluctuation. In this situation, the collapse
of the attractive Bose gas is opposed by the thermal energy(\cite{8}  $3\hbar\omega [\frac{kT}{\hbar\omega}]^4 \zeta(4)\sim\hbar\omega N^{4/3}$) of the
atoms and consequently, we get the
estimation of the critical number as
$\frac{N_ca}{l}\sim[\frac{l}{a}]^{1/2}>1$.

However, close to the condensation temperature, the length scale of the
system becomes\cite{8} $L_{T}\sim l\sqrt{\frac{kT}{\hbar\omega}}$.
For this reason the estimation of the critical number would be
$\frac{N_ca}{l}\sim [\frac{l}{a}]^5\gg 1$.

In this paper we will explicitly calculate the critical numbers by a
scaling theory within the Hartree-Fock(H-F) approximation and will
explicitly show its temperature dependence. Before doing that, let
us see how the critical number for $T=0$, was calculated by the
scaling theory of Baym and Pethick \cite{14}.

\section{Zero temperature scaling theory for the collapse of BEC}
For $T\rightarrow 0$, the
system of $N$ weakly interacting Bose particles is well described by
the Gross-Pitaevskii(G-P) equation
 \bea
i\hbar\frac{\partial \Psi_0({\bf{r}},\tau)}{\partial\tau}=
(-\frac{\hbar^2\nabla^2}{2m}+V({\bf{r}})+g\mid\Psi_0({\bf{r}},\tau)\mid^2)\Psi_0({\bf{r}},\tau),
 \eea
where $V({\bf{r}})=\frac{1}{2}m\omega^2r^2$ is the harmonic
potential and $\Psi_0({\bf{r}},\tau)$ is the order parameter for
Bose condensate and is a function of space({\bf{r}}) and
time($\tau$). A stationary solution of the G-P equation have the
form
$\Psi_0({\bf{r}},\tau)=\Psi_0({\bf{r}})e^{-\frac{i\mu\tau}{\hbar}}$,
where $\mu$ is the chemical potential. The stationary solution of
this form can be obtained by minimizing the energy
functional\cite{8}
 \bea
E=\int(\frac{\hbar^2}{2m}\mid\nabla\Psi_0({\bf{r}})\mid^2+V({\bf{r}})\mid\Psi_0({\bf{r}})\mid^2+\frac{g}{2}\mid\Psi_0({\bf{r}})\mid^4)d^3{\bf{r}}
 \eea for a fixed number of
particles($N=\int\mid\Psi_0\mid^2d{\bf{r}}$). The stationary
solution without current gives
$\Psi_0({\bf{r}})=\sqrt{n({\bf{r}})}$, where $n({\bf{r}})$ is the
number density of particles at the position ${\bf{r}}$. For
stationary and for no current condition, we have the energy
functional from the Eqn.(2) as
 \bea E(n)=\int
d{\bf{r}}[\frac{\hbar^2}{2m}\mid\nabla\sqrt{n}\mid^2+nV+\frac{g}{2}n^2].
 \eea
For the noninteracting gas($g=0$), the  solution of G-P equation
must be the ground state of harmonic oscillators. Hence, for $g=0$,
the solution of the G-P equation is
$\sqrt{\frac{N}{l^3\pi^{3/2}}}e^{-\frac{r^2}{2l^2}}$. For $g<0$, the
solution of G-P equation was obtained by a scaling
theory(variational approach) based on the Gaussian
function\cite{10,14}. From this scaling theory we can write
$\Psi_0({\bf{r}})=\sqrt{n}=\sqrt{\frac{N}{\nu^3l^3\pi^{3/2}}}e^{-\frac{r^2}{2\nu^2l^2}}$,
where $\nu$ is a scaling parameter which fixes the width of the
condensate. Since the attractive interaction causes the reduction of
the width of the trap, $\nu$ must be less than one. With this
solution, the energy functional of the Eqn.(3) can be written in
units of $N\hbar\omega$ as \cite{14}
 \bea
X_0(\nu)=[E(\nu)/N\hbar\omega]=[\frac{3}{4}(1/\nu^2+\nu^2)-\frac{Na}{\sqrt{2\pi}l}\frac{1}{\nu^3}].
 \eea
According to FIG. 1, this energy functional has a minimum and a
maximum below a critical number of particles($N_c$). In equilibrium,
the system must be at the minimum of the energy functional. About
this minimum, $\nu$ has a stable value. Above the critical number of
particles, $E(\nu)$ has no stable minimum. In this situation, the
energy of the system arbitrarily reduces to take $\nu=0$ and the gas
collapses to vanish its width. $N_c$ is calculated by requiring that
the first and second derivative vanish at the critical point. From
this scaling theory we can have\cite{14,10,8}
$\frac{N_ca}{l}=0.671$. According to the numerical result\cite{21}
$\frac{N_ca}{l}=0.575$ and according to the experimental
result\cite{20} $\frac{N_ca}{l}=0.459$. However, the same problem
was also attacked by many authors \cite{22,23,24,25}. Kagan, Surkov
and Shlyapnikov discussed the dynamic properties of a trapped
Bose-condensed gas under variations of the confining field and found
analytical scaling solutions for the evolving condensate\cite{22}.
Pitaevskii addressed the condition of collapse from the dynamical
point of view\cite{23}. Yukalov and Yukalova calculated the optimal
trap shape allowing for the condensation of the maximal number of
atoms with negative scattering lengths\cite{24}. Savage, Robins and
Hope solved the Gross-Pitaevskii equation numerically for the
collapse, induced by a switch from positive to negative scattering
lengths\cite{25}.  Although the G-P equation is extensively used for
the Bose gas with attractive interaction, yet it is to be noted
that, it was rigorously derived for repulsive interaction\cite{26}.

\begin{figure}
\includegraphics{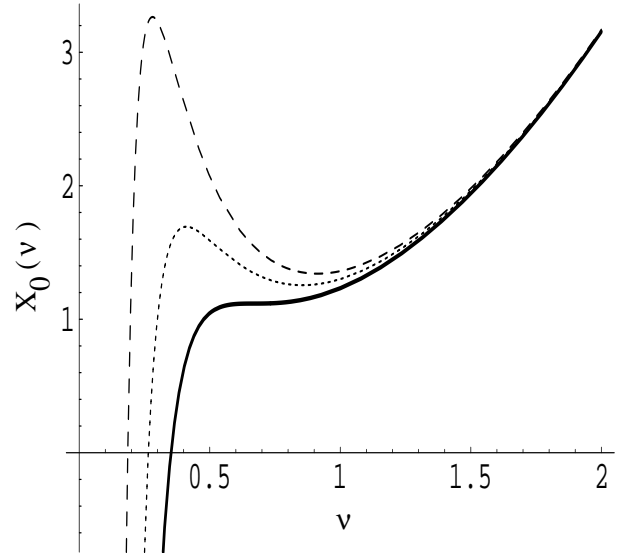}
\caption {Plot of energy per particle ($X_0(\nu)$)in units
$\hbar\omega$ with the scaling parameter $\nu$. For the thick line,
$\frac{Na}{l}=0.671$. For the dotted line, $\frac{Na}{l}=0.5$. For
the dashed line, $\frac{Na}{l}=0.35$. All the lines follow from the
Eqn.(4).} \label{fig:Energy per particle at zero temperature}
\end{figure}

\section{Scaling assumption for the collapse of BEC at finite temperature}
Although the problem of an attractive Bose gas at finite temperature
was discussed by many authors\cite{27,28,29}, yet the temperature
dependence of critical number of particles with a scaling theory has
not been calculated. The authors of the Refs.\cite{27,28} considered
the collapse of the condensate only. The authors of the
Ref.\cite{29} explored the collapse for $T=0$ and for $T\gg T_c$.
But we are considering the collapse for $0\le T\le T_c$ and are
considering the fact that both the condensed cloud and the thermal
cloud of the Bose gas would collapse due to the attractive type of
interaction. In the fifth section we will calculate $N_c$ by the
following scaling theory within the Hartree-Fock(H-F) approximation
which was discussed extensively for Bose gas in the Ref.\cite{8}.

For the interacting Bose gas, average occupation numbers $\bar{n}_i$
as well as single particle wave functions $\phi_i$ are
determined\cite{8} by minimizing the grand potential
$\Omega=(E-TS-\mu N)$, where $S=k\sum_i
(1+n_i)ln(1+n_i)-(n_i)ln(n_i)$ is the entropy and $i$ runs from $0$
to $\infty$. The entropy($S$) and the total number of
particles($N=\sum_in_i)$ both are the functions of $\{n_i\}$.
Minimizing the grand potential with respect to the occupation
number($n_i$) we get the average occupation number as
$\bar{n}_i=\frac{1}{e^{(\epsilon_i-\mu)/kT}-1}$. Since the entropy
and the total number of particles depend on the occupation numbers
$\{n_i\}$, the single particle wave functions $\{\phi_i\}$ are
simply obtained by minimizing the energy functional with proper
normalization constraint $\int d^3{\bf{r}}\mid\phi_i\mid^2=1$
\cite{8}. Within the Hartree-Fock approximation, we have the
expression of energy functional as\cite{8}
\begin{eqnarray}
E&=&\int d^3{\bf{r}}[\frac{\hbar^2}{2m}n_0\mid \nabla\phi_0\mid^2+\sum_{i\neq 0}\frac{\hbar^2}{2m}n_i\mid \nabla\phi_i\mid^2
\nonumber\\&&+V(r)n_0({\bf{r}})+V({\bf{r}})n_T({\bf{r}})+\frac{g}{2}n_0^2({\bf{r}})\nonumber\\&&+2gn_0({\bf{r}})n_T({\bf{r}})+gn_T^2({\bf{r}})]
\end{eqnarray}
Minimizing the above energy functional of the Eqn.(5) we have the
following equations\cite{8}. \bea [-\frac{\hbar^2}{2m}\nabla^2
+V({\bf{r}})+g(n_0({\bf{r}})+2n_T({\bf{r}}))]\phi_0=\mu\phi_0 \eea

 \bea [-\frac{\hbar^2}{2m}\nabla^2
+V({\bf{r}})+2gn({\bf{r}})]\phi_i=\epsilon_i\phi_i
 \eea
From the Eqns.(6) and (7) we can determine the wave functions
$\{\phi_i\}$ and the energy eigenvalues $\{\epsilon_i\}$ for a
given $n_0({\bf{r}})$ and $n_T({\bf{r}})$. In our scaling theory we
introduce a scaling parameters $\nu$, which would fix the width of
the condensed and thermal clouds as well as would take proper
choices of $n_0({\bf{r}})$ and $n_T({\bf{r}})$. Eventually the
choices of  the wave functions $\{\phi_i\}$ are denoted by the
scaling parameter $\nu$. Since the minimum of the grand potential
corresponds to the equilibrium of the system, the minimum of the
energy functional for a certain choice of $\{\phi_i\}$ would
correspond to the equilibrium of the system. So, for equilibrium
condition of the system, the energy functional would be minimized
with respect to the scaling parameter($\nu$). In the FIG. 2, we
will see that above a critical number of particles($N_c$), the
energy functional has no stable minimum. If the energy functional as
well as the grand potential has no stable minimum, the grand
potential as well as the energy functional would arbitrarily
decrease until the width of the system becomes zero. Under this
condition the system is said to be collapsed.

In absence of the interaction, the condensed cloud is described by
the wave function
$\phi_0({\bf{r}})=\frac{1}{\sqrt{l^3\pi^{3/2}}}e^{-r^2/2l^2}$ and by
the density $n_0({\bf{r}})=\frac{n_0}{l^3\pi^{3/2}}e^{-r^2/l^2}$. In
absence of interaction, the number density of the excited particles
is\cite{8}
$n_T({\bf{r}})=\int_0^{\infty}\frac{1}{e^{(\frac{p^2}{2m}+\frac{m\omega^2r^2}{2})/kT}-1}\frac{4\pi
p^2dp}{(2\pi\hbar)^3}=\frac{1}{\lambda_T^3}g_{\frac{3}{2}}(e^{-\frac{m\omega^2r^2}{2kT}})$,
where $\lambda_T=\sqrt{\frac{2\pi\hbar^2}{mkT}}$, and
$g_{\frac{3}{2}}(x)=x+x^2/2^{3/2}+x^3/3^{3/2}+...$ is Bose-Einstein
function of a real variable $x$. The length scale of the excited
particles in thermal equilibrium is
$L_T\sim\sqrt{\frac{kT}{m\omega^2}}=l\times\sqrt{\frac{kT}{\hbar\omega}}$.
Since the length scale($L_T$) of the thermal cloud is proportional
to the length scale($l=\sqrt{\hbar/m\omega}$) of the condensed
cloud, the two length scale would reduce by the same factor due to
the attractive interaction.

In presence of the attractive interaction the wave function of the
condensed would become
 \bea
\phi_0(r)=\frac{1}{\sqrt{\nu^3 l^3\pi^{3/2}}}e^{-r^2 /2 \nu^2 l^2}
 \eea
and the density of the thermal cloud would become
 \bea
n_T({\bf{r}})=\frac{1}{\nu^3
\lambda_T^3}g_{\frac{3}{2}}(e^{-\frac{m\omega^2r^2}{2\nu^2 kT}}).
 \eea
The same scaled form of $n_T({\bf{r}})$ is also obtained from the
variation of B-E statistics such that
$\bar{n}({\bf{p}},{\bf{r}})=\frac{1}{e^{(\frac{p^2\nu^2}{2m}+\frac{m\omega^2r^2}{2\nu^2})/kT}-1}$.
With this variational form of statistics we have the total number of
excited particles as $N_T=\int d^3{\bf{r}}\frac{1}{\nu^3
\lambda_T^3}g_{\frac{3}{2}}(e^{-\frac{m\omega^2r^2}{2\nu^2
kT}})=(\frac{kT}{\hbar\omega})^3\zeta(3)$. For $T=T_c$, there is no
particles in the ground state and we have
$N=(\frac{kT_c}{\hbar\omega})^3\zeta(3)$. For $T\le T_c$, the total
number of particles in the ground state is
$n_0=N-N_T=N(1-\frac{T}{T_c})^3$.

\section{Evaluation of the Hartree-Fock energy functional}

Let us now evaluate the seven terms of energy functional in the
Eqn.(5). Since we have scaled the length and there is a square of the
derivative of the wave functions with respect to the length, all the
kinetic energy terms of Eqn.(5) would be proportional to the inverse
squared of the scaling parameter. Due to the presence of
the term quadratic in length in the expression of harmonic potential
energy, all the potential energy terms of Eqn.(5) would be
proportional to the squared of the scaling parameter. Similarly, as
the interaction energy goes like the inverse of the volume, all the
interacting energy terms of Eqn.(5) would be proportional to the
inverse cubed of the scaling parameter.

The first term of the energy functional is the kinetic
energy of the condensed cloud and is obtained as
\begin{eqnarray}
&&\frac{\hbar^2}{2m}\frac{n_0}{\nu^3l^3\pi^{3/2}}\frac{4\pi}{(\nu^2\l^2)^2}\int_{0}^{\infty}r^2e^{-r^2/\nu^2l^2}r^2dr
\nonumber\\&&=\frac{3}{4}N\hbar\omega\frac{1}{\nu^2}[1-(\frac{T}{T_c})^3].
\end{eqnarray}
The second term of the energy functional is the kinetic energy of
the thermal cloud and is approximately obtained as
\begin{eqnarray}
&&\int_{0}^{\infty}\int_{0}^{\infty}\frac{p^2}{2m}\frac{1}{e^{(\frac{p^2\nu^2}{2m}+\frac{m\omega^2r^2}{2\nu^2})/kT}-1}\frac{4\pi
p^2dp}{(2\pi\hbar)^3}4\pi
r^2dr\nonumber\\&&=\frac{3}{2}\frac{1}{\nu^2}(\frac{N}{\zeta(3)})^{4/3}(\frac{T}{T_c})^4\zeta(4)\hbar\omega.
\end{eqnarray}
The third term of the energy functional is the potential energy of
the condensed cloud and is obtained as
\begin{eqnarray}
&&\int_0^{\infty}\frac{1}{2}m\omega^2r^2
[\sqrt{\frac{n_0}{\nu^3l^3\pi^{3/2}}}e^{-r^2/2\nu^2l^2}]^24\pi
r^2dr\nonumber\\&&=\frac{3}{4}\nu^2\hbar\omega
N[1-(\frac{T}{T_c})^3].
\end{eqnarray}
The fourth term of the energy functional is the potential energy of
the thermal cloud and is obtained as
\begin{eqnarray}
&&\frac{1}{\nu^3\lambda_T^3}\int_0^{\infty}[g_{\frac{3}{2}}(e^{-\frac{m\omega^2r^2}{2\nu^2kT}})]\frac{1}{2}m\omega^2r^24\pi
r^2dr\nonumber\\&&=\frac{3}{2}\nu^2(\frac{N}{\zeta(3)})^{4/3}(\frac{T}{T_c})^4\zeta(4)\hbar\omega.
\end{eqnarray}
The fifth term of the energy functional is the interaction energy of
the particles solely in the condensed cloud and is obtained as
\begin{eqnarray}
&&\frac{2\pi\hbar^2 a}{m}\frac{n_0^2}
{(\nu^3l^3\pi^{3/2})^{2}}\int_0^{\infty}e^{-\frac{2r^2}{\nu^2l^2}}4\pi
r^2
dr\nonumber\\&&=\frac{N^2}{\sqrt{2\pi}}\frac{1}{\nu^3}\frac{a}{l}\hbar\omega[1-(\frac{T}{T_c})^3]^2.
\end{eqnarray}
The sixth term of the energy functional is the interaction energy of
the particles in the condensed and thermal clouds and is obtained as
\begin{eqnarray}
&&\frac{32\pi^2\hbar^2a}{m}\frac{n_0}{\nu^3l^3\pi^{3/2}}\frac{1}{\nu^3\lambda_T^3}\int_{0}^{\infty}e^{-\frac{r^2}{\nu^2l^2}}g_{\frac{3}{2}}
(e^{-\frac{m\omega^2r^2}{2\nu^2kT}})r^2
dr\nonumber\\&&=\frac{8\pi\hbar^2an_0}{m\nu^6\lambda_T^3l^3}\sum_{i=1}^{\infty}(\frac{2kT\nu^2
l^2}{i^2m\omega^2
l^2+i2kT})^{3/2}\nonumber\\&&\approx\frac{2^{3/2}\sqrt{\zeta(3/2)}}
{\sqrt{\pi}\nu^{3}}\frac{a}{l}N^{3/2}(\frac{T}{T_c})^{3/2}[1-(\frac{T}{T_c})^3]\hbar\omega.
\end{eqnarray}
The seventh term of the energy functional is the interaction energy
of the particles solely in the thermal cloud and is obtained as
\begin{eqnarray}
&&\frac{4\pi\hbar^2
a}{m}\frac{4\pi}{\nu^6\lambda_T^6}\sum_{i,j=1}^{\infty}\int_{0}^{\infty}\frac{e^{-\frac{m\omega^2}{2\nu^2kT}(i+j)r^2}r^2}{i^{3/2}j^{3/2}}dr
\nonumber\\&&=\sqrt{\frac{2}{\pi}}\frac{S'N^{3/2}}{\nu^3[\zeta(3)]^{3/2}}\frac{a}{l}\hbar\omega[\frac{T}{T_c}]^{9/2}
\end{eqnarray} where $S'=\sum_{i,j=1}^{\infty}\frac{1}{(ij)^{3/2}(i+j)^{3/2}}\approx
0.6534$.

The energy functional($E$) of the Eqn.(5) is now a function of the
variational parameter $\nu$. Above the critical number of particles,
$E(\nu)$ has the lowest value($-\infty$) at $\nu=0$. So, above the
critical number of particles, the width of the system would be zero
to signalize the collapse. Let us define the energy per particle in
units of $\hbar\omega$ as $X_t(\nu)=\frac{E(\nu)}{N\hbar\omega}$.
Let us also write $\frac{T}{T_c}=t$. Now collecting all the terms of
the energy functional from the Eqn.(10) to (16), we have
\begin{eqnarray}
X_t(\nu)&=&\frac{3}{4}(\frac{1}{\nu^2}+\nu^2)[1-t^3]+\frac{3}{2}\frac{N^{1/3}\zeta(4)}{[\zeta(3)]^{4/3}}t^4(\frac{1}{\nu^2}+\nu^2)
\nonumber\\&-&[\frac{1}{\sqrt{2\pi}}\frac{Na}{l}[1-t^3]^2+\sqrt{\frac{8\zeta(3/2)}{\pi}}\frac{N^{1/2}a}{l}t^{3/2}[1-t^3]
\nonumber\\&&+S'\sqrt{\frac{2}{\pi[\zeta(3)]^3}}\frac{N^{1/2}a}{l}t^{9/2}]\frac{1}{\nu^3}\nonumber\\&=&c_1(t)[\frac{1}{\nu^2}+\nu^2]-\frac{c_2(t)}{\nu^3}
\end{eqnarray}
where
 \bea
c_1(t)=\frac{3}{4}(1-t^3)+\frac{3}{2}\frac{N^{1/3}\zeta(4)
t^4}{[\zeta(3)]^{4/3}}
 \eea
and
\begin{eqnarray}
c_2(t)&=&\frac{1}{\sqrt{2\pi}}\frac{Na}{l}[1-t^3]^2+\sqrt{\frac{8\zeta(3/2)}{\pi}}\frac{N^{1/2}a}{l}t^{3/2}[1-t^3]\nonumber\\&+&S'\sqrt{\frac{2}{\pi[\zeta(3)]^3}}
\frac{N^{1/2}a}{l}t^{9/2}.
\end{eqnarray}
From the Eqns.(18) and (29), and putting $\frac{a}{l}=0.0066$
\cite{20} and $t=0.8$, in the Eqn.(17) we can write
\begin{eqnarray}
X_{0.8}(\nu)&=&(0.366+0.5203N^{1/3})(\frac{1}{\nu^2}+\nu^2)\nonumber\\&&-\frac{0.0069N^{1/2}+0.000627N}{\nu^3}
\end{eqnarray}
The Eqn.(20) represents the FIG. 2. In the FIG. 2 we see that below
a critical number of particles, $X_t(\nu)$ has a stable minimum and
an unstable maximum. At the critical number of particles, the minimum
and maximum coincide. Above the critical number of particles, there
is no stable minimum and the energy of the system arbitrarily decreases to
$-\infty$ and its width($\nu l$) becomes zero to signalize the
collapse of the gas.

\begin{figure}
\includegraphics{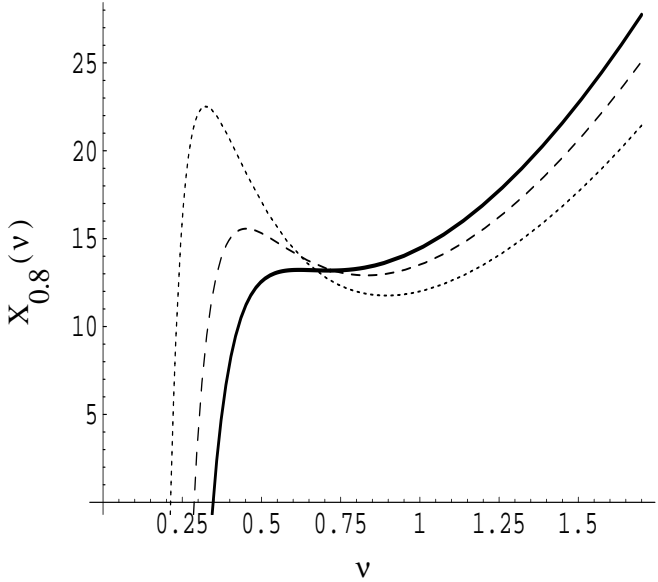}
\caption {Plot of energy per particle ($X_{0.8}(\nu)$)in units
$\hbar\omega$ with the scaling parameter $\nu$. For the thick line,
$\frac{Na}{l}=27.78$. For the dashed line, $\frac{Na}{l}=20$. For
the dotted line, $\frac{Na}{l}=12$. All the lines follow from the
Eqn.(20).} \label{fig:Plot of Energy functional}
\end{figure}
\section{Critical conditions for the collapse}

For a minimum of $X_t(\nu)$ as well as for the equilibrium condition
of the system, we must have $\frac{\partial X_t}{\partial\nu}=0$. At
the critical point($\nu_c$), the minimum and maximum coincide. So,
at the critical point, we must have $\frac{\partial^2
X_t}{\partial\nu^2}\mid_{\nu_c}=0$. For the critical point($\nu_c$),
the Eqn.(17) gives following equations
 \bea \frac{\partial
X_t}{\partial\nu}\mid_{\nu_c}=-c_1(t)\frac{2}{\nu^3}+2c_1(t)\nu+c_2(t)\frac{3}{\nu^4}=0
\eea and \bea \frac{\partial^2
X_t}{\partial\nu^2}\mid_{\nu_c}=c_1(t)\frac{6}{\nu^4}+2c_1(t)-c_2(t)\frac{12}{\nu^5}=0.
 \eea
From the Eqns.(21) and (22) we have $\nu_c=(1/5)^{1/4}=0.66874$ and
$c_2(t)=\frac{8}{15}\nu_cc_1(t)$, which represents the equation for
the critical number($N_c$) and can be recast as
\begin{eqnarray}
&&\frac{N_ca}{\sqrt{2\pi}l}[1-t^3]^2+\sqrt{\frac{8\zeta(3/2)}{\pi}}\frac{N_c^{1/2}a}{l}t^{3/2}[1-t^3]
\nonumber\\&&+S'\sqrt{\frac{2}{\pi[\zeta(3)]^3}}\frac{N_c^{1/2}a}{l}t^{9/2}=\frac{8}{15}\nu_c[\frac{3}{4}(1-t^3)
\nonumber\\&&+\frac{3}{2}\frac{N_c^{1/3}\zeta(4)t^4}{[\zeta(3)]^{4/3}}].
\end{eqnarray}

From the Eqn.(23) we get the expression of $\frac{N_ca}{l}$ for
$t=0$ as
\begin{eqnarray}
\frac{N_ca}{l}=\frac{\sqrt{2\pi}}{2}[\frac{4\nu_c}{5}]=0.671
\end{eqnarray}
Now we see our scaling theory for $T>0$ to be consisted with that
given by Baym and Pethick for $T=0$ \cite{14}.

For $0<t<1$, we can consider the second and third terms of the left
hand side and the first term of the right hand side of the Eqn.(23)
as the perturbing terms. We expand its left hand side with respect to 
its first term and similarly, we expand its right hand side with respect 
to its second term. Now we get the critical number from the Eq.(23) up to three leading orders in $l/a$ as
\begin{eqnarray}
\frac{N_ca}{l}&=&\frac{[\sqrt{2\pi}\zeta(4)]^{3/2}}{[\zeta(3)]^2}[\frac{4\nu_c}{5}]^{3/2}[\frac{l}{a}]^{1/2}\frac{t^6}{[1-t^3]^3}
\nonumber\\&&+\frac{3\sqrt{2\pi}[\frac{4\nu_c}{5}]}{4[1-t^3]}\nonumber\\&&-\frac{3}{2}[\sqrt{\frac{8\zeta(3/2)}{\pi}}t^{3/2}(1-t^3)
+S'\sqrt{\frac{2}{\pi[\zeta(3)]^3}}t^{9/2}]\nonumber\\&&\times\frac{[2\pi]^{7/8}[\zeta(4)]^{3/4}}{[\zeta(3)]}[\frac{4\nu_c}{5}]^{3/4}
[\frac{a}{l}]^{1/4}\frac{t^3}{[1-t^3]^{7/2}}\nonumber\\&=&1.210[\frac{l}{a}]^{1/2}\frac{t^6}{(1-t^3)^3}
+\frac{1.006}{[1-t^3]}-[10.666
t^{3/2}\nonumber\\&&\times(1-t^3)+1.636t^{9/2}][\frac{a}{l}]^{1/4}\frac{t^3}{[1-t^3]^{7/2}}.
\end{eqnarray}
The Eqn.(25) represents the behavior of the critical number with
temperature. For\cite{20} $\frac{a}{l}=0.0066$, the change of
critical number($N_c$) with $t$ is shown in FIG. 3. From this figure
and according to the Eqn.(25), we get $\frac{N_ca}{l}=27.89$ for
$t=0.8$. The Eqn.(25) represent an approximate relation between
$N_c$ and $t$. However, from the FIG. 2, we see that
$\frac{N_ca}{l}=27.78$ for $t=0.8$.
\begin{figure}
\includegraphics{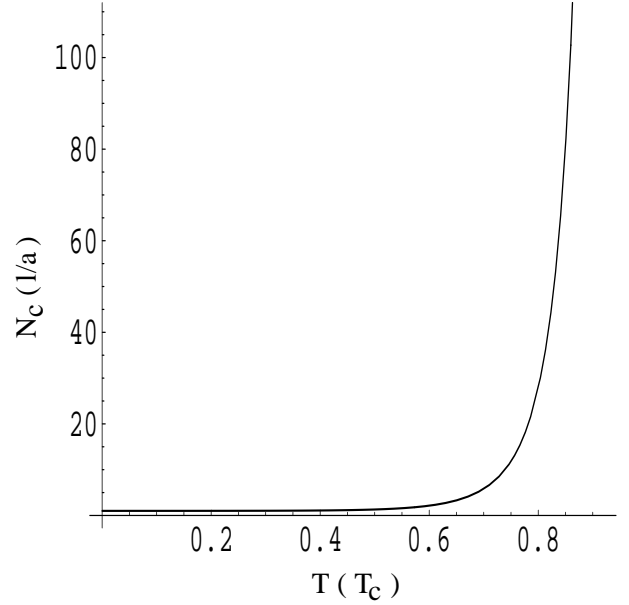}
\caption {Plot of the critical number $N_c$ in units of
$\frac{l}{a}$ with temperature($T$) in units of ($T_c$). This plot
follows form the Eqn.(25).} \label{fig:Critical Number of Particles
with Temperature}
\end{figure}

At the condensation temperature i.e. as $t\rightarrow 1$, only the
third term of the left hand side of the Eqn.(23) contributes to give
\begin{eqnarray}
\frac{N_ca}{l}&=&[\frac{l}{a}]^5[\frac{\pi}{2}]^3\frac{\zeta(3)[\zeta(4)]^6}{S'^6}[\frac{4\nu_c}{5}]^6\nonumber\\&=&2.253[\frac{l}{a}]^5
\end{eqnarray}
From the above equation we see that near the condensation
temperature $N_c^{1/6}\frac{a}{l}$(=1.145) is of the order of unity
and is also expected from the qualitative point of view.

\section{Conclusions}

Initially we explained the physics of the collapse of the attractive
atomic Bose gas. Then we qualitatively estimated $N_ca/l$ for the
various range of temperatures ($0\le T\le T_c$). Finally we
calculate $N_ca/l$ by a scaling theory within the Hartree-Fock
approximation. Our calculation supports the qualitative estimation
of $N_ca/l$.

In the FIG. 3, we see that the temperature dependence of $N_c$ is
not significant for $0\le T\lesssim 0.5 T_c$. If experiment of
collapse is performed within this range of temperature, the zero
temperature theory would well satisfy the experimental result.

Here we consider an attractive type of delta potential which is not
an actual potential but a model potential \cite{14,8,10}. The
problem with this model potential is that, no bound state is formed
for a finite number of particles in a three dimensional delta
potential. So, for a true delta potential, the collapse of a finite
number of particles is impossible. However, once we do the scaling
assumption, the collapse is not impossible for a delta potential and
we can calculate the critical numbers.

It is a coincidence of our scaling theory that the grand potential
and the energy are minimized simultaneously for a certain choice of
the scaling parameter. However, in general, the grand potential and
the energy are not minimized simultaneously. If we want to calculate
the critical number by some other theory, the minimization of the
energy may not be sufficient for the calculation of the critical
number. In that case, some modification would come from the entropy
and chemical potential.

Although the Hartree-Fock approximations are extensively used\cite{13,8} even for the critical region($T_c$) of Bose systems, yet the Hartree-Fock-Bogoliubov(H-F-B) approximation is a more reliable one\cite{28}. Any mean-field approximation(H-F, H-F-B, etc.) is also known to be less suitable for the critical region.

We considered the interaction as
$V_{int}({\bf{r}})=g\delta^3({\bf{r}})$. For repulsive interaction,
the density of the gas in the central region decreases. This would
cause a slight decrease in $T_c$ as shown theoretically in the
Ref.\cite{13} and experimentally in the Refs.\cite{7,30}. On the
other hand, for attractive interaction, the density in the central
region of the trap would increase and we expect slight increase in
$T_c$. However, our scaling theory does not predict any $T_c$ shift.
This is a limitation of our scaling theory. We should look for a
better theory which would simultaneously predict the critical
numbers and the $T_c$ shift.

\section{Acknowledgment}

Several useful discussions with J. K. Bhattacharjee of SNBNCBS, Deepak Dhar of TIFR and Koushik Ray of IACS are gratefully acknowledged. We also acknowledge the hospitality of the Centre for Nonlinear Studies in Hong Kong Baptist University.

\end{document}